\documentclass[11pt,a4paper]{article}
\usepackage{jcappub}

\title{The fluxes of CN neutrinos from the Sun in case of mixing in a spherical layer in the solar core}
\author[a]{Anatoly Kopylov,}
\author[a]{Valery Petukhov}
\affiliation[a]{Institute of Nuclear Research of Russian Academy of Sciences \\
117312 Moscow, Prospect of 60th Anniversary of October Revolution
7A, Russia}
\emailAdd{kopylov@inr.ru}
\emailAdd{beril@inr.ru}

\abstract{The results of the calculation are presented for the
fluxes of CN neutrinos from the Sun in case of mixing in a spherical
layer in the solar core, consistent with the seismic data and with
the measured solar neutrino fluxes. It is shown that a substantial
increase of the flux of $^{13}N$ neutrinos can be gained in this
case. The possible implications for experiment are discussed.}

\arxivnumber{1110.5703}

\begin{document}

\maketitle

\section{Introduction.}

Solar neutrinos and helioseismology are complimentary tools in
studying solar structure. The measured fluxes of solar neutrinos
\cite{1} - \cite{7} and the found parameters of neutrino
oscillations \cite{8} are in a good agreement with the predictions
of the standard solar model \cite{9}. However, still there is "solar
abundance problem" which represents the fact that best available
photospheric abundances of high Z elements are at odds with the
helioseismic measurements \cite{10}. This stimulates the search for
some modifications of the solar model.

Because few acoustic modes penetrate to the nuclear active region
where the solar energy is generated, there is still very scares
experimental data for the region 0.1 -- 0.2 $R_{\odot}$ which could
illuminate the dynamics of the solar core \cite{11}. In the article
\cite{12} the attempt has been made to set the stringent limit for
the average mean molecular weight in the inner 20\% of the Sun by
radius $\mu_c$ (mean molecular weight of the solar core). In other
paper \cite{13} it was suggested that the variation in the sound
speed in the solar core might be also caused by a partial mixing. In
the framework of a new approach (Linear Solar Model \cite{14}) the
constraints on opacity (and composition) were discussed in
\cite{15}. Unfortunately, there are still no direct experimental
determinations of these quantities and it would be very interesting
to get some direct experimental data for the solar core. The flux of
$^{13}N$ neutrinos is very sensitive to the tiny variations of the
abundance of $^{12}C$ across the solar core while the flux of
$^{15}O$ neutrinos is practically stable by any variation of
abundances. This might be the key for solving the problem of the
dynamics of the solar core emphasizing the measured ratio of the
fluxes of $^{13}N$ and $^{15}O$ neutrinos as the parameter directly
characterizing the mixing in the solar core.  The increase of
accuracy in the measurements of the effect from $\nu-e^{-}$
scattering in electronic detectors promises in the future the
precise measurement of this effect not only beyond the upper bound
of $^7Be$ neutrinos, but also beyond the upper bound of $^{13}N$
neutrinos, thus facilitating the finding of this ratio. This would
be the next step in the solar neutrino research.

In this article we follow previous works \cite{16}, \cite{17} where
the calculation of the effect of mixing on the fluxes of solar
neutrinos were presented for the mixing within all central zone. It
was shown that a substantial increase of the flux of $^{13}N$
neutrinos can be obtained in this case while this model with mixing
still is consistent with the seismic data and the measured solar
neutrino fluxes. This effect is due to the large gradient of
concentrations of $^{12}C$ on solar radius so that even very mild
mixing can increase substantially the abundance of $^{12}C$ in a
nuclear active region, where $^{13}N$ is produced. Here the question
is addressed what effect can be obtained if the mixing occurs only
in a spherical layer within the radiative zone. Two parameters are
used for the layer: the radius of the bottom of the layer $R_m$ and
the depth $\Delta R_m$ in units of Solar radius so that the top of
the layer is at the radius $R_m+ \Delta R_m$. The mixing is
conceived as a mild process running during a time interval, short in
comparison with the age of the Sun. The effect of this mixing would
be the homogeneous distribution of all elements within full
spherical layer while the temperature gradient still is preserved of
the one of the standard solar model. The limited geometry of the
mixing in this case should not change drastically the parameters
fixed by helioseismology while the effect on the flux of neutrinos
from the decay of $^{13}N$ can be large while all other fluxes of
solar neutrinos will be left practically unchanged. This is the main
point of our approach.

It is worth to note that mixing changes the profile of the mean
molecular weight along the radius of the Sun  and consequently it
changes also the opacity profile. This, in turn, affects the whole
solar structure, as it was shown in \cite{15}. However, as it was
noted, for example, in \cite{18} so far there are no direct seismic
determinations of these quantities in the region of the solar core.
The most precise parameter characterizing this region is a mean
molecular weight of the solar core. This value has been found in
\cite{12} with the uncertainty 1\% at the level of 95.4\% ($2
\sigma$) C.L.  Thus, the parameters of mixing should not result in
the change of the mean molecular weight of the solar core exceeding
1\% in comparison with the standard model. It is also worth to note
that the effects produced by variations of opacity in distinct zones
of the Sun may compensate each other, as it was noted in \cite{15}.
In our case the change of the mean molecular weight along the radius
of the Sun has different signs: in lower layer of the mixing zone
$\delta \mu < 0$ and in upper one: $\delta \mu > 0$. Thus, the
changes of opacity in these two layers due to the changes of the
mean molecular weight are partially compensated. We did not go into
details of how the sound speed profile will be changed after mixing
limiting ourselves only by the mean molecular weight of the solar
core values. Being the most sensitive probe provided by
helioseismology the sound speed profile in case of mixing demands a
further study. The detailed consideration of this question should
probably be addressed somewhere else as a next step of our approach.

\section{Calculation.}

Let's suggest that during a time interval, short in comparison with
the age of the Sun, the solar matter in a certain layer $\Delta R_m$
has got mixed. As a result of this mixing the abundance of all
elements has been averaged within all this layer (Figure 1).

\begin{figure}[!ht]
\centering
\includegraphics[width=4in]{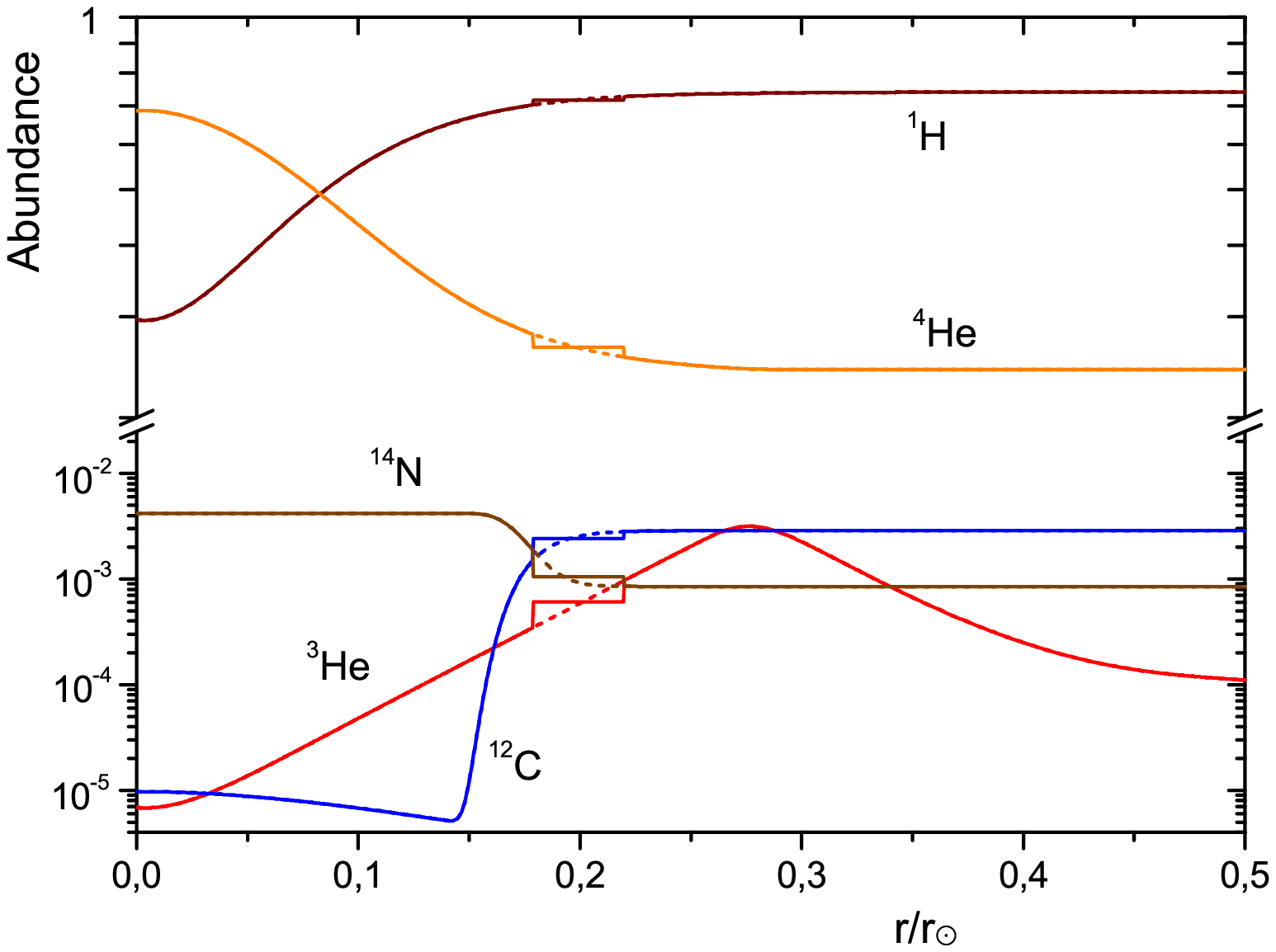}
\caption{The profile for the abundance of $^1H$, $^3He$, $^4He$,
$^{12}C$ and $^{14}N$ before (dashed line) and after (solid line)
mixing.}
\end{figure}

\noindent Let's also suggest that profile of temperature and density
across the Sun has not been changed due to the partial compensation
of the changes of $\delta \mu$ in the lower and upper layers of the
mixed zone. If we divide the matter of the Sun on thin spherical
layers so that inside these layers the temperature and the density
can be taken to be constant, then the evolution of the abundance of
the elements in each layer can be described by the set of
differential equations (\cite{17}).

{\footnotesize
\begin{equation}
\label{eq1} \left\{ {\begin{array}{l}
dX_1/dt =-X_1^2 \frac{3}{2} (\alpha _{11} +\alpha _{11}^{'})+X_3^2 \frac{1}{9} \alpha _{33} -X_3 X_4 \frac{1}{12} \alpha _{34} \\
dX_3/dt =X_1^2 \frac{3}{2} (\alpha _{11} +\alpha _{11}^{'})-X_3^2 \frac{1}{3} \alpha _{33} -X_3 X_4 \frac{1}{4} \alpha _{34} \\
dX_4/dt =X_3^2 \frac{2}{9} \alpha _{33} +X_{3} X_{4}\frac{1}{3}\alpha _{34} \\
dX_{12}/dt =-X_1 X_{12} \alpha _{112} +X_1 X_{15} \frac{12}{15} \alpha _{115} \\
dX_{13}/dt =-X_1 X_{13} \alpha _{113} +X_1 X_{12} \frac{13}{12} \alpha _{112} \\
dX_{14}/dt =-X_1 X_{14} \alpha _{114} +X_1 X_{13} \frac{14}{13} \alpha _{113} \\
dX_{15}/dt =-X_1 X_{15} \alpha _{115} +X_1 X_{14} \frac{15}{14} \alpha _{114} \\
 \end{array}} \right.
\end{equation}}

\noindent Here the first three equations describe the variations of
the abundance of elements  $X_i$ of the atomic weight $i$ in the
reactions of  {\it pp}-chain ($^1H$, $^3He$ and $^4He$), and the
latter four -- in the reactions of {\it CNO}-cycle ($^{12}C$,
$^{13}C$, $^{14}N$ and $^{15}N$). Here $\alpha_{ij} = \rho f_{ij}$ ,
where $f_{ij}$ -- the rates of the reaction $A_i+A_j$ from \cite{19}
with the known S-factor \cite{20}, \cite{21} ($\alpha _{11}^{'}$
corresponds to the reaction $^1H+^1H+e^-$), $\rho$ -- the density in
the given layer. The small contribution in the first equation of the
reactions of the type $X_1 X_{15} \alpha _{115}$ were neglected, as
well as the processes of gravitational settling and diffusion.

\begin{figure}[!ht]
\centering
\includegraphics[width=3in]{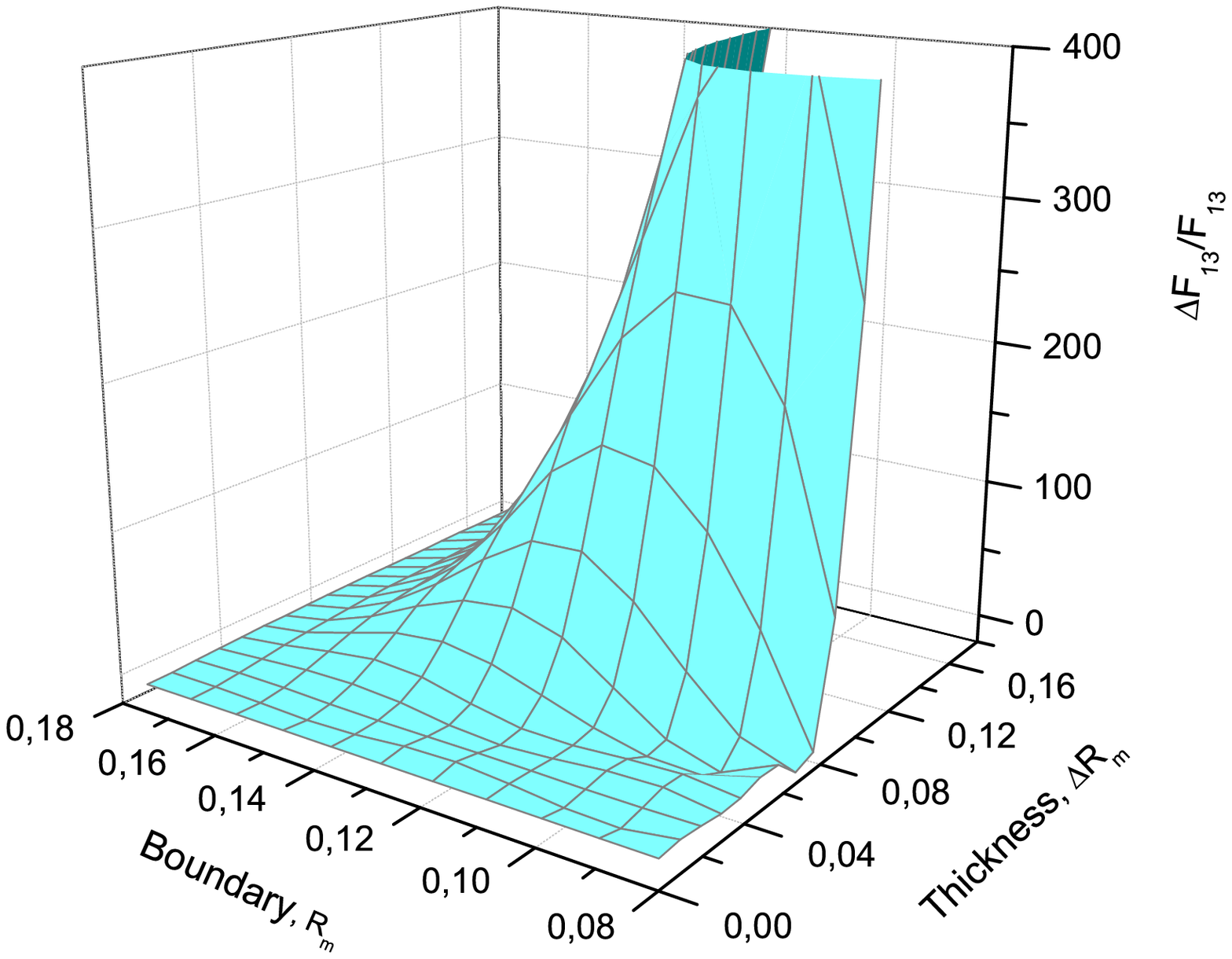}
\includegraphics[width=3in]{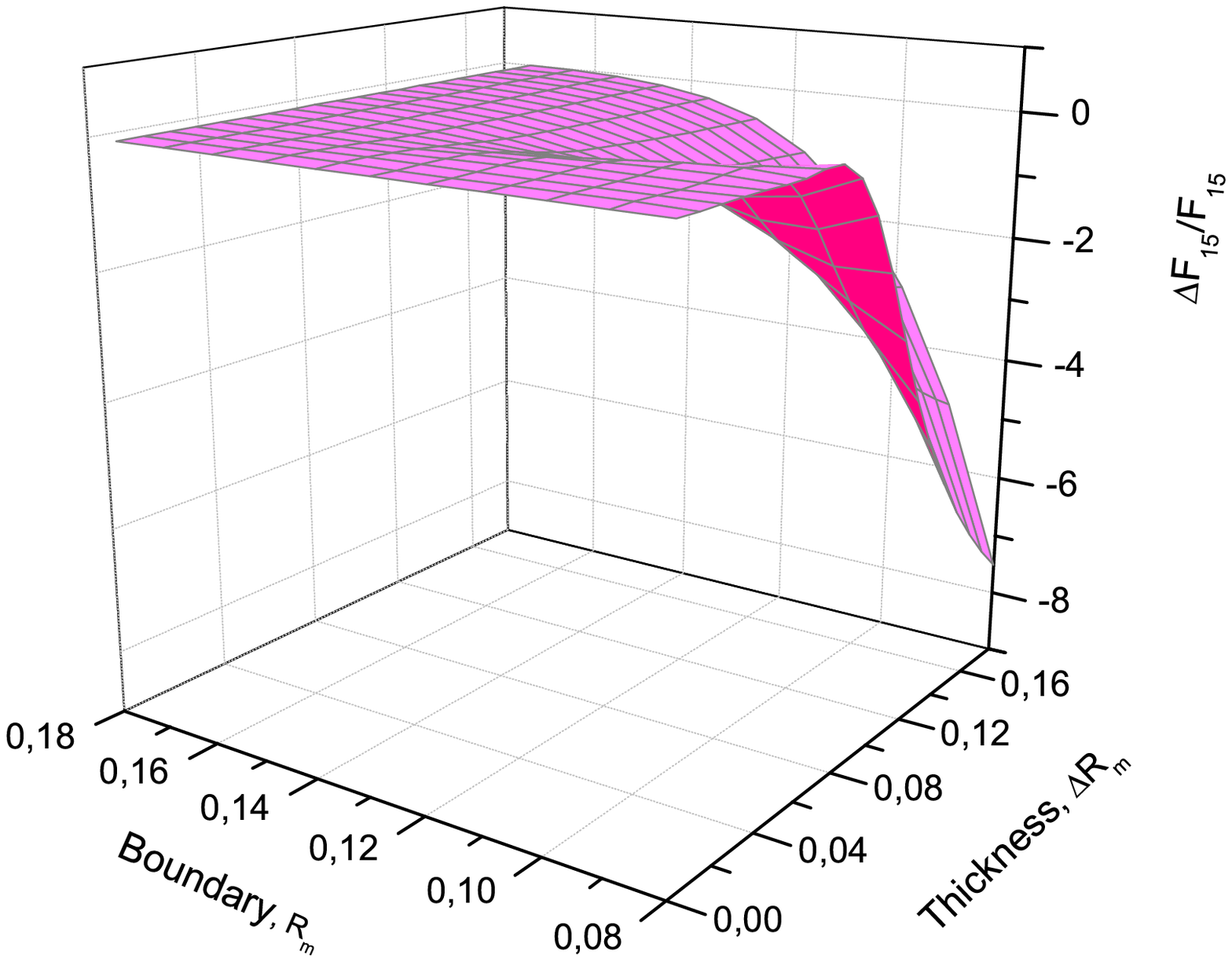}
\includegraphics[width=3in]{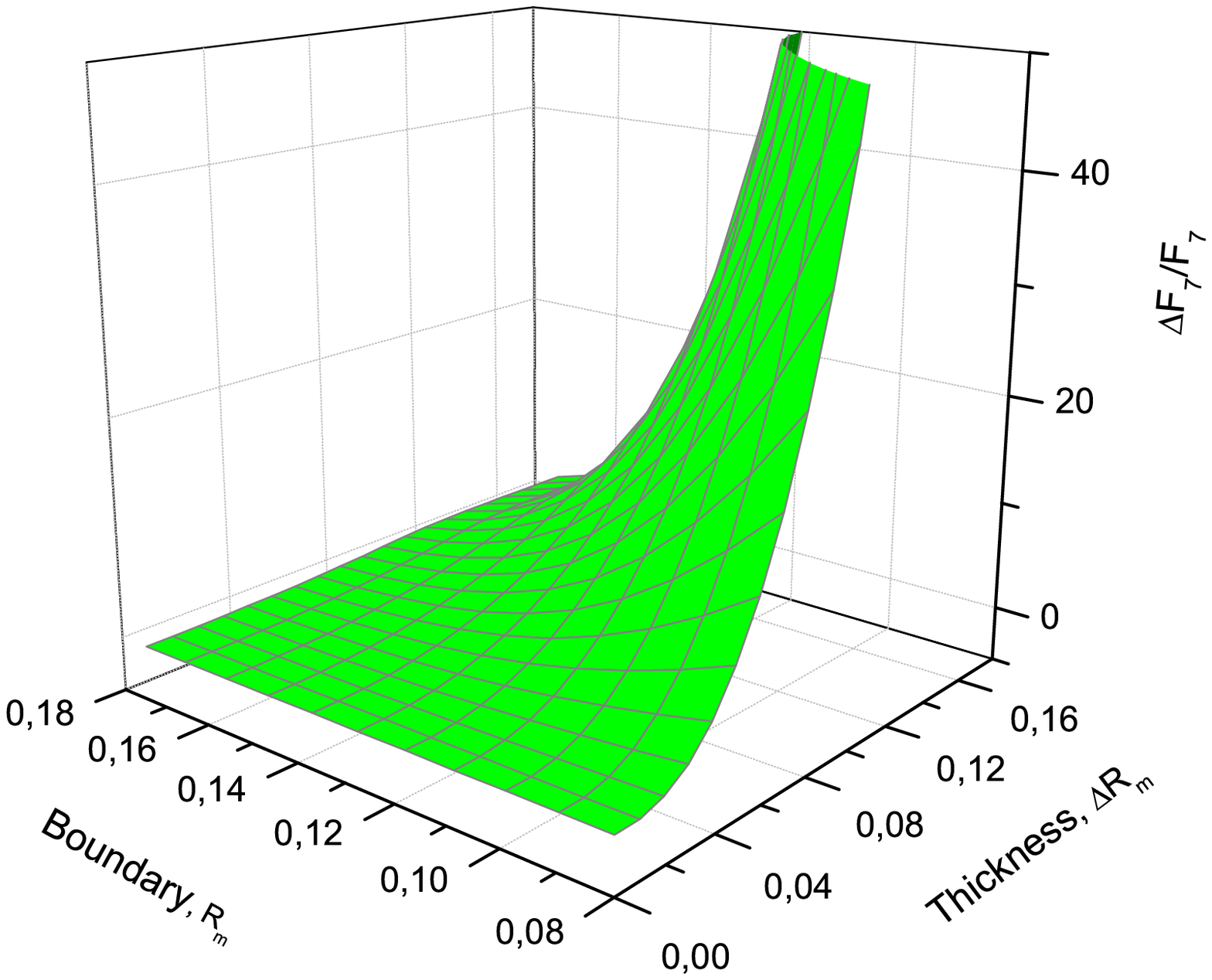}
\includegraphics[width=3in]{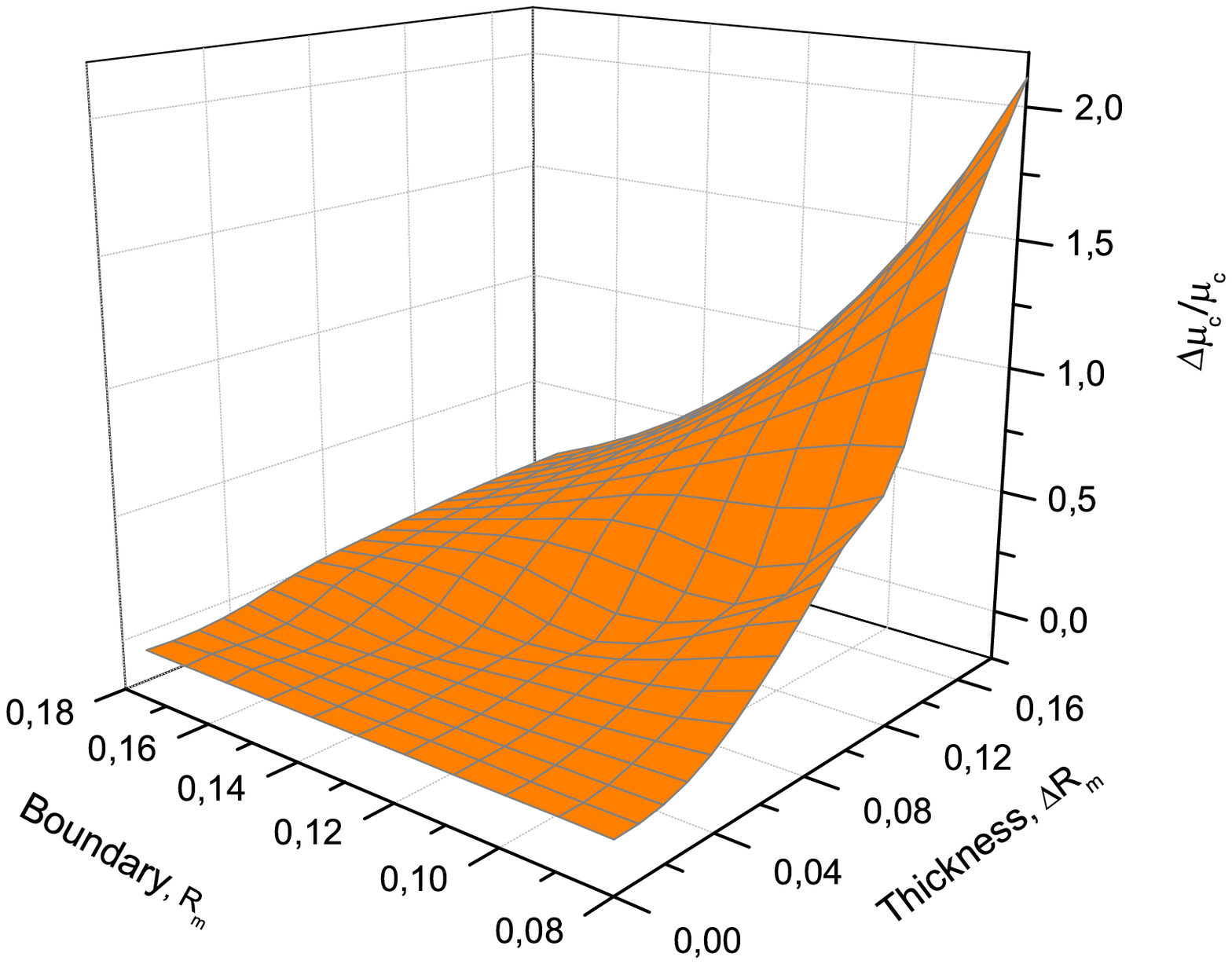}
\caption{Relative changes of the neutrino fluxes $F_{13}$, $F_{15}$,
$F_7$ and of the mean molecular weight of the solar core $\mu_c$ (in
{\%}) by mixing.}
\end{figure}

Similar to \cite{16}, the set \eqref{eq1} was resolved by 4th order
Runge-Kutt method. As the initial abundance of elements the present
one observed on the surface of the Sun has been used. The time
interval has been taken equal to the age of the Sun (4.6 billion
years). Thus, by resolving this set of equations we obtain the
abundance of elements in each layer at the present time. In the same
way the abundance of elements and the corresponding fluxes of solar
neutrinos are calculated in case of mixing. The total flux of
neutrino from the Sun is determined by the reaction rates and the
abundance of elements. For the flux of neutrinos from the decay of
$^{13}N$ :

\begin{equation} \label{eq2_}
F_{13}(t)=4 \pi N_{A} R_{\odot }^3 \int \limits_0^1 {\rho (r)X_1
(r,t)\frac{X_{12}(r,t)}{12} \alpha _{112} (r)dr}
\end{equation}

\begin{figure}[!ht]
\centering
\includegraphics[width=4in]{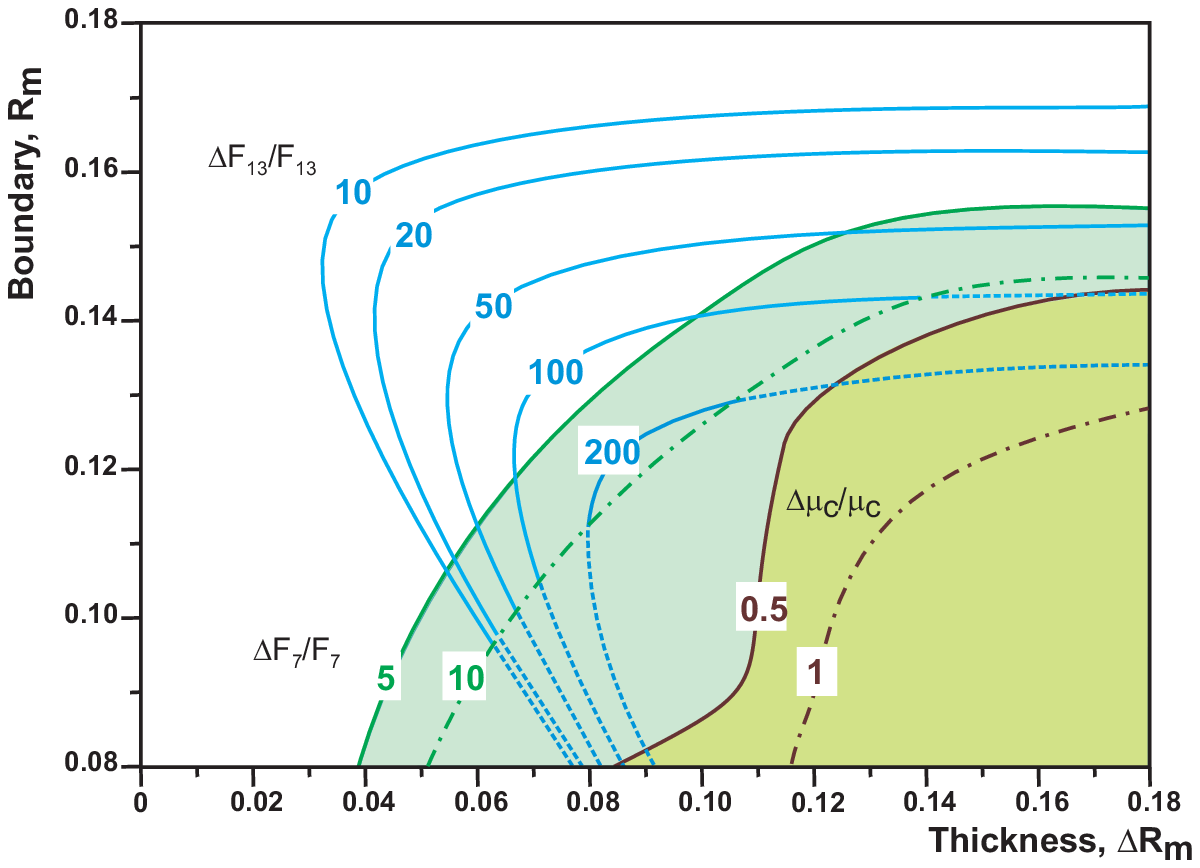}
\caption{Contours for the relative changes of the neutrino fluxes
$F_{13}$, $F_{15}$, $F_7$ and of the mean molecular weight of the
solar core $\mu_c$ (in {\%}) by mixing within the layer. The painted
regions correspond to the changes of  $\mu_c$ and of the flux of
beryllium neutrino greater than 1$\sigma $ (0.5\% and 5\%
correspondingly). Lines marked by line-dots correspond to 2$\sigma
$.}
\end{figure}

\noindent Similar expressions can be written for the neutrino fluxes
$F_{15}$ and $F_7$ from the decay of $^{15}O$ and $^7Be$. The
neutrino fluxes will be changed by mixing in a certain layer with
the lower boundary $R_m$ and the thickness $\Delta R_m$. The
parameters fixed by helioseismology will be changed also,
particularly, the mean molecular weight of the solar core $\mu_c $
\cite{12}.

We have calculated these values before and after mixing which
happened at certain time. The resulting relative changes of the
neutrino fluxes and the mean molecular weight of the solar core
$\mu_c$ are presented on Figure 2. At Figure 3 the results obtained
are presented as contours (with the exclusion of $F_{15}$ which has
been changed marginally in comparison with $F_{13}$).

\begin{figure}[!ht]
\centering
\includegraphics[width=3in]{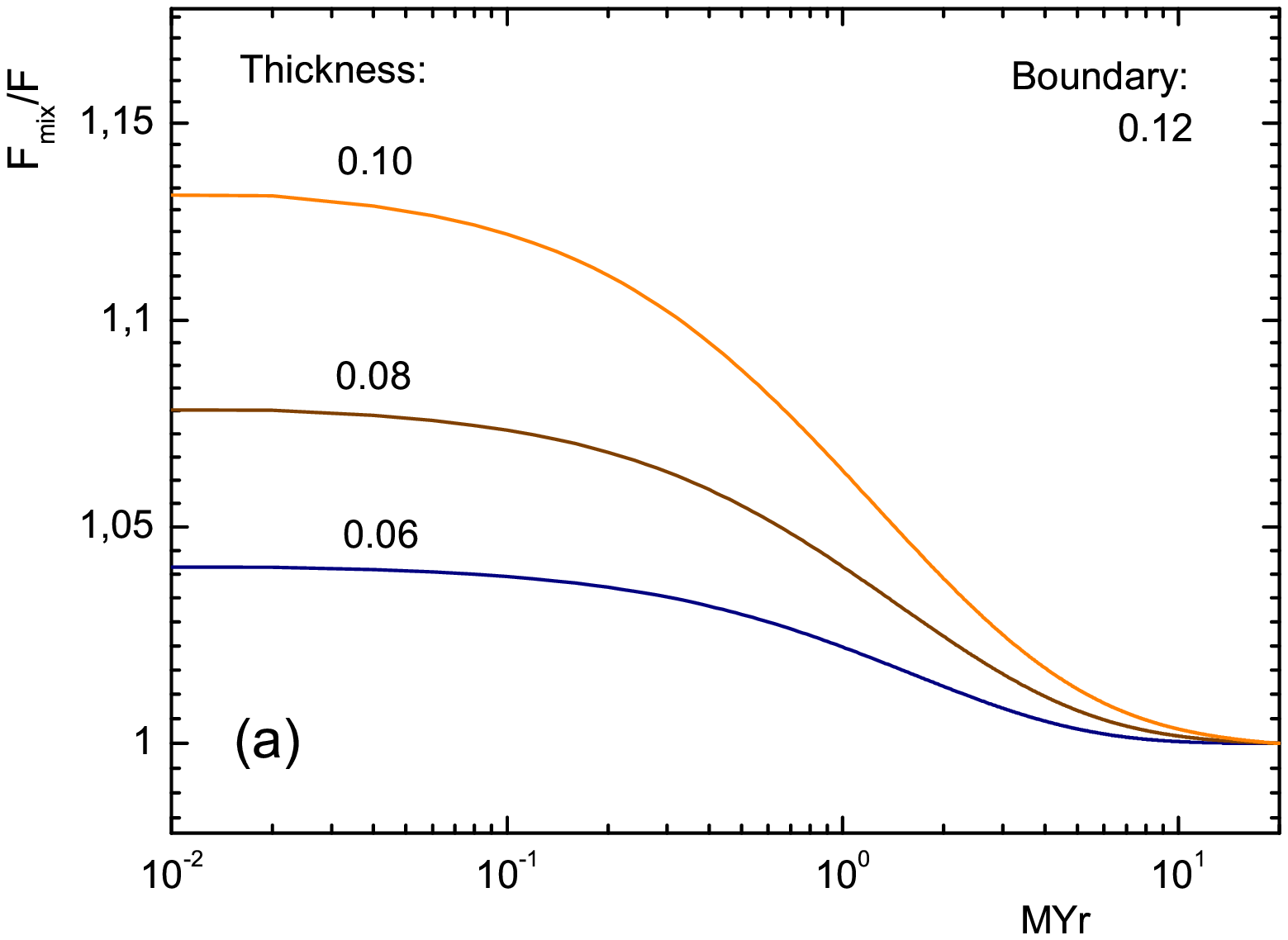}
\includegraphics[width=3in]{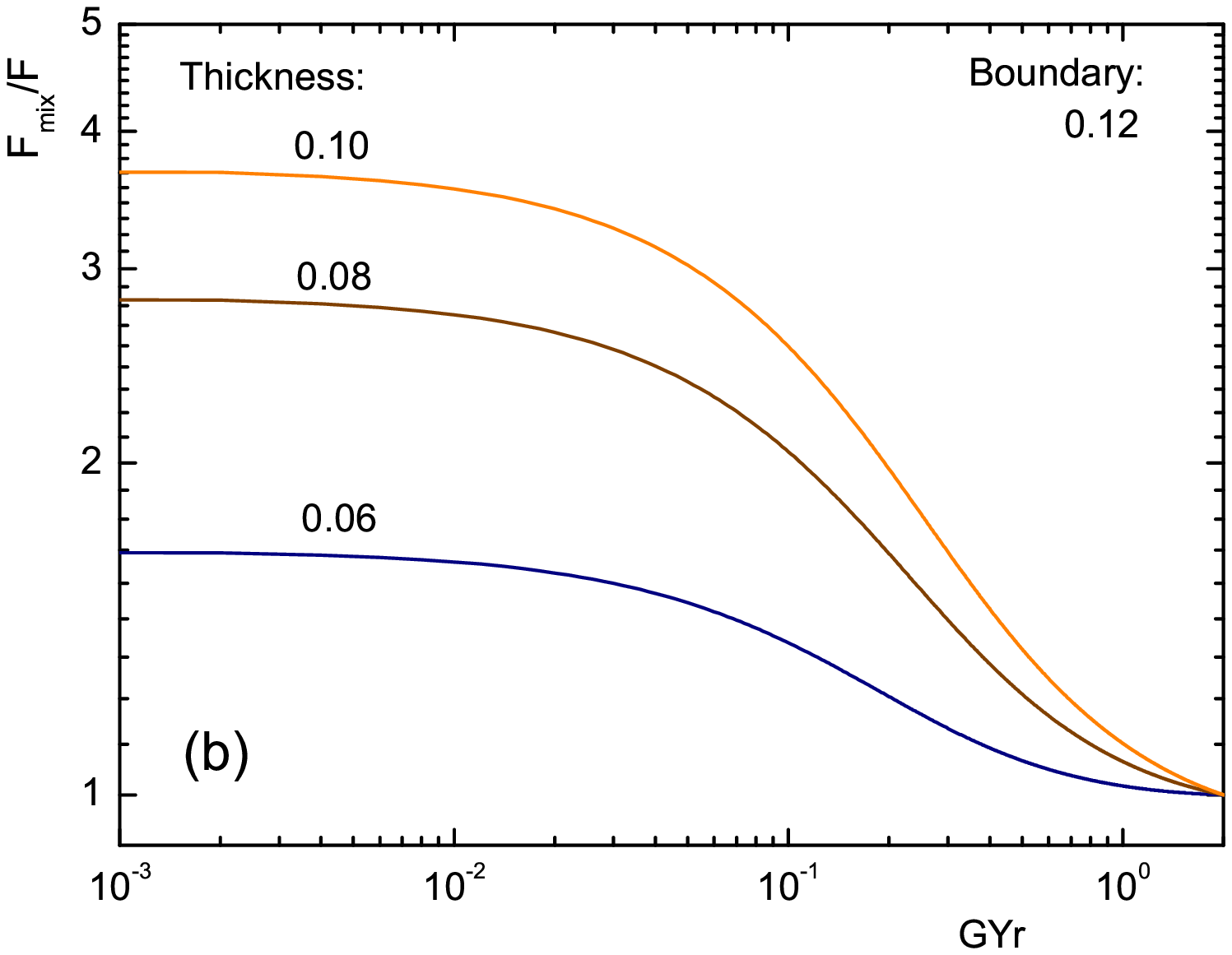}
\caption{The evolution of the flux of (a) $^7Be$- and (b)
$^{13}N$-neutrino after mixing (the boundary of the mixed layer
0.12, the thickness of the layer 0.06, 0.08 and 0.1 of the radius of
the Sun.}
\end{figure}

One can see that the allowed region for the parameters used is
mainly determined by the uncertainty in the determination of the
flux of beryllium neutrinos (5\% in Borexino experiment \cite{5}).
Let's look how the neutrino fluxes are changed after mixing in a
layer. Figure 4 shows the evolution of the fluxes of beryllium and
$^{13}N$-neutrino in case of mixing ($F_{mix}$ -- the flux after
mixing in a layer, $F$ -- the flux without mixing at the same
moment). Figure 4 shows, that the flux of beryllium neutrino
recovers more quickly the stationary value than the flux of
$^{13}N$-neutrino. This is explained by higher rate of the reaction
$\rm{^3He}+\rm{^4He} \rightarrow \rm{^7Be}+\gamma$ (determining the
neutrino generation in the reaction $\rm{^7Be}+e^- \rightarrow
\rm{^7Li}+\nu _e$) than the one of the reaction $\rm{^{12}C}+p
\rightarrow \rm{^{13}N}+\gamma$ (determining the neutrino generation
in the reaction$\rm{^{13}N} \rightarrow \rm{^{13}C}+e^++\nu
_e$).Therefore beryllium neutrino flux returns to the stationary
value faster then the flux of $^{13}N$-neutrino. Let's see how the
picture presented on Figure 3 is changed in a certain time, for
example, in 10 million years. The results are presented on Figure 5.
We see that the allowed change of the flux of $^{13}N$-neutrino is
limited only by the mean molecular weight of the solar core fixed by
helioseismology and may exceed a few hundreds percent.

\begin{figure}[!ht]
\centering
\includegraphics[width=4in]{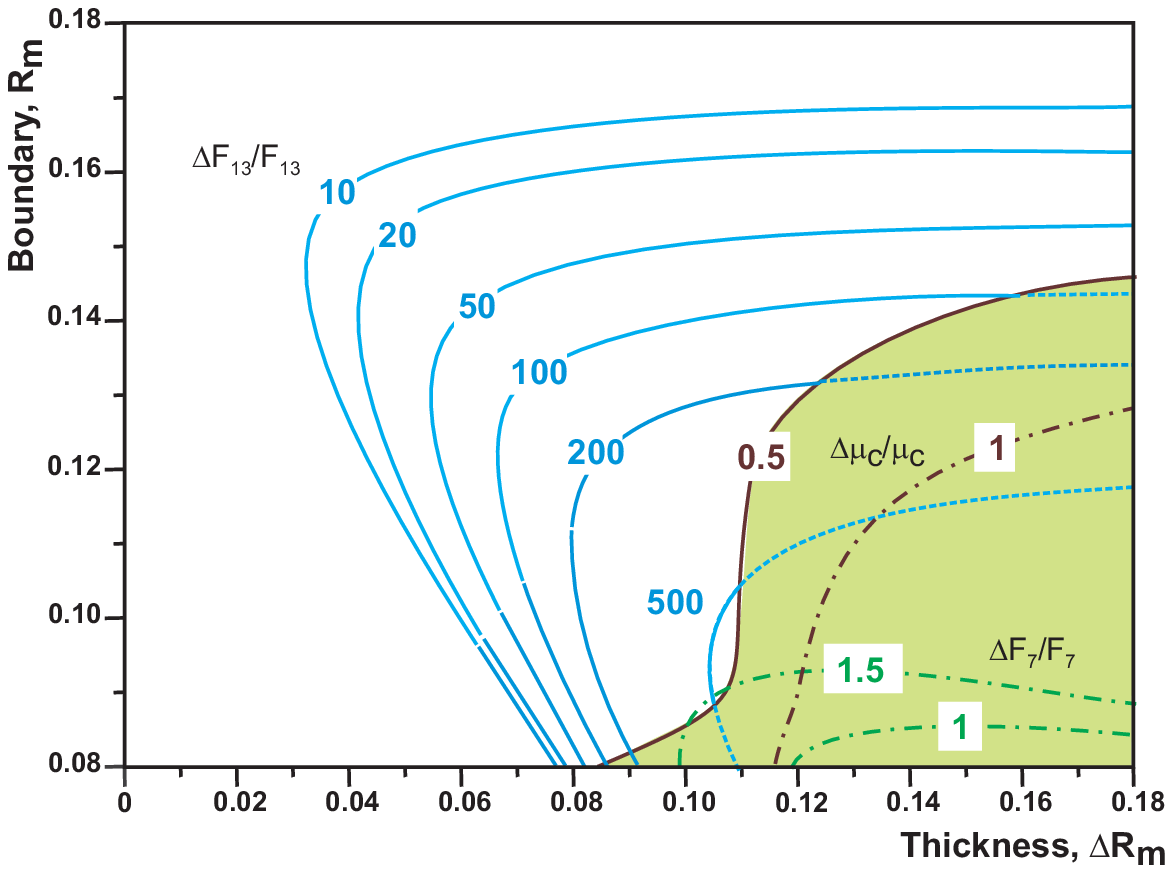}
\caption{Contours for the relative changes of the neutrino fluxes
$F_{13}$, $F_{15}$, $F_7$ and of the mean molecular weight of the
solar core $\mu_c$ (in {\%}) 10 million years after mixing.}
\end{figure}

A recent publication of Borexino experiment \cite{6} gives an upper
limit for CNO neutrino flux $7.7 \cdot 10^8$ $cm^{-2}s^{-1}$ what
corresponds to less then $1.5$ for the ratio between Borexino
measurements and predictions of the High Metallicity (GS98) SSM
\cite{10}. Because the energy of the end-point of $^{13}N$-neutrino
is lower than the one of $^{15}O$-neutrino the weight of
$^{13}N$-neutrino in the total flux of CNO neutrino in Borexino
experiment is approximately 5 times lower than the weight of
$^{15}O$-neutrino \cite{22}. Thus, the change of the flux of
$^{13}N$-neutrino on the level of a few hundreds percent still can
be considered as compatible with this latest results of Borexino
experiment.

\section{Conclusions.}

The results of the calculation show that there is the possibility of
mixing in a spherical layer compatible with the measured fluxes of
solar neutrinos, with the parameters fixed by helioseismology and at
the same time producing substantial increase of $^{13}N$ neutrinos.
The flux of $^{15}O$ neutrinos is changed only marginally in
comparison with the predicted by standard solar model. In this model
the ratio of the fluxes of $^{13}N$ and $^{15}O$ neutrinos is equal
to 1.40 with the uncertainty of about 0.05 for different parameters
of solar model, see, for example, Table 3 of \cite{10}. As one can
see from Figure 5 the mixing can bring this ratio to many sigmas
above this value. This constitutes the clear signature of mixing in
the solar core and presents the motivation for future experiment
provided it is capable to measure both neutrino fluxes of $^{13}N$
and $^{15}O$ neutrinos. The increase of accuracy in the measurements
of the effect from  $\nu-e^{-}$ scattering in electronic detectors
promises in the future the precise measurement of the effect not
only beyond the upper bound of $^7Be$ neutrinos, but also beyond the
upper bound of $^{13}N$ neutrinos, thus facilitating the finding of
this ratio. This would be the next step in the solar neutrino
research.

\section{Acknowledgements.}
We warmly acknowledge funding from the Programs of support of
leading schools of Russia (grant  {\#}3517.2010.2.) and we
appreciate a very substantial commentary of referee which was very
helpful in the improvement of the first version of this article.

\end{document}